\documentclass[a4paper,11pt]{article}
\pdfoutput=1 % if your are submitting a pdflatex (i.e. if you have
\usepackage{jinstpub} % for details on the use of the package, please
\title{Gas gap studies about streamer operated RPCs.}

\author[1]{A. Paoloni,\note{Corresponding author.}}
\author{A. Mengucci,}
\author{M. Spinetti,}
\author{M. Ventura,}
\author{and L. Votano.}
\affiliation{INFN-LNF, \\via E. Fermi 40 00044 Frascati (RM), Italy.}

\emailAdd{alessandro.paoloni@lnf.infn.it}

\abstract{
The requirement of high rate capability for operation at LHC, led 
20 years ago to the achievement of Resistive Plate Chambers operated in
avalanche mode, thanks to the introduction of new gas mixtures and to the
development of the Front-End electronics.
The need for a further increase of the rate capability, in view of the
upgrades of LHC, is imposing new detector geometries with thinner gas gaps
and electrodes.
Streamer operation of RPCs may still be suitable for low rate experiments,
and therefore in this paper a comparison between two different detector 
geometries, the old standard and the newly proposed one, is performed in 
streamer mode.
}

\keywords{Gaseous detectors, Resistive plate chambers, Electrical discharge in gases.}

\begin{document}
\maketitle
\flushbottom

\section{Introduction}
\label{sec:intro}
Resistive Plate Chambers (RPCs) are gaseous detectors used in LHC experiments 
\cite{blhc} because of their high rate capability and time resolution in 
avalanche mode.
Operated in streamer mode, they have been also used in large area apparatuses 
for neutrino \cite{bopera} and cosmic ray \cite{bargo} physics.

Present studies about bakelite electrode RPCs are mainly addressed to the 
improvement of the rate capability in avalanche mode for the upgrades of ATLAS 
and CMS experiments at LHC, involving the reduction 
of the gas gap and of the electrode thickness \cite{bATLASup}.  

In view of a possible use of single gap streamer-operated RPCs at low 
rates, a comparison has been done between the detector performances with
two geometries: 

\begin{itemize} 
\item geometry 1: 2 mm thick gas gap and 2 mm thick bakelite electrodes (so far
the standard in use);
\item geometry 2: 1 mm thick gas gap and 1.2 mm thick bakelite electrodes.
\end{itemize} 

All the chambers used in the tests have been produced by the General Tecnica,
where RPCs have been produced for many experiments, such as ATLAS, CMS, ALICE,
BABAR, ARGO and OPERA.

In section \ref{sec:setup} the experimental set-up is briefly described,
together with the analysis.
The results are displayed in section \ref{sec:results}, in terms of 
efficiency, multi-streamer probability, timing properties and streamer
signal parameters.
A summary of the main conclusions is given in section \ref{sec:conc}.

\section{Experimental set-up and analysis description}
\label{sec:setup}
The experimental set-up is composed by five RPCs, $10 \times 10$ cm$^2$ wide,
piled up.
Three chambers, made according to geometry 2, are used to trigger cosmic rays.
In between them, there are two other RPCs under test, one similar to the
trigger ones and the other one produced according to geometry 1.
All the chambers are read-out by means of copper pads covering the whole
detector surface.
Negative polarity induced signals are discriminated at 100 mV (150 mV) for
1 mm (2 mm) gas gap RPCs and sampled by means of a CAEN N6742 digitizer 
\cite{bcaen} at 5 GS/s.
Pedestals are evaluated on a single waveform basis, averaging the first 200 
samples.

\begin{figure}
\centering
\includegraphics[width=\textwidth]{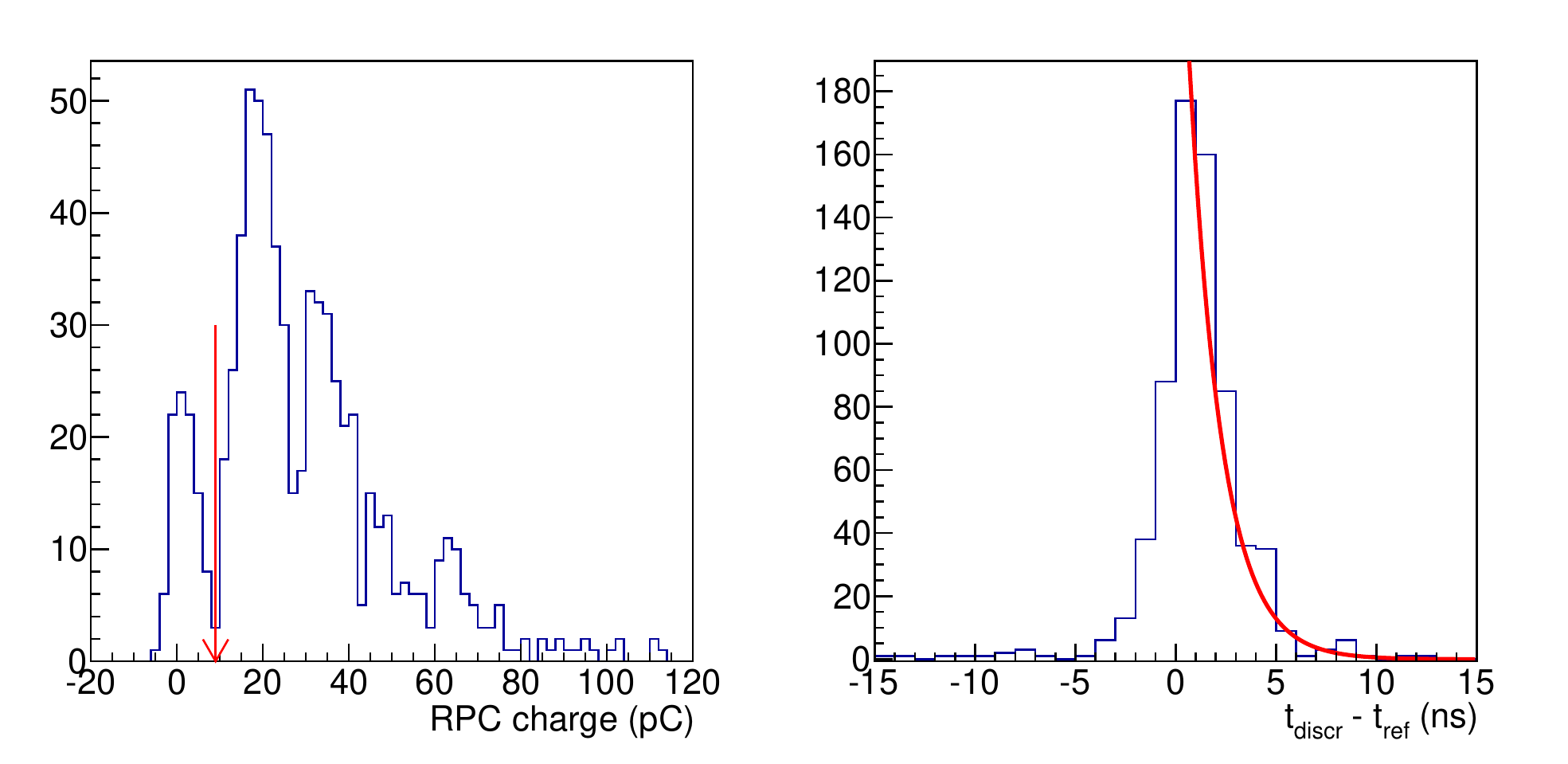}
\caption{Examples of prompt charge (left) and streamer formation time (right)
distributions. The streamer formation time is measured with respect to one of
the trigger RPCs.}
\label{fig0}
\end{figure}

The efficiency is measured using the prompt charge distribution,
obtained by integrating the read-out signal waveform.
The detector is considered efficient if the measured charge is greater than
a cut value defined in such a way to separate the events without streamers, 
as shown in figure \ref{fig0}. 
The obtained values are compatible within statistical errors with the
efficiencies measured by counting the events with the signal waveform 
crossing the discrimination threshold.

The multi-streamer probability is estimated checking by eye a sub-sample of
the acquired events; multi-streamer events are defined as those with either
after-pulsing or streamer signals with very high amplitude (twice that of
standard events).
In a similar way a sample of events is selected in order to measure the single
streamer parameters: prompt induced charge, amplitude, rise-time 
(from 10\% to 90\% of the amplitude) and Full Width Half Maximum (FWHM).

The timing parameters, i.e. the streamer formation time and the detector
time resolution are measured from the time difference between the discriminated
signals of the RPCs under test and one of the trigger chambers, used as a 
reference ($t_{discr} - t_{ref}$).
The time of the maximum is used as estimator for the difference between the
streamer formation times of the RPC under test and of that used as a
reference.
The right queue of the distribution is fitted with an exponential and the
time constant is taken as an estimation of the time resolution.
An example of the distribution with the fit is shown in figure \ref{fig0}.

The induced signal amplitude is influenced both by the charge ionized in the 
gas (dependent on the gap) and also by the ratio between the electrode 
thickness and the gas gap.
Indeed, according to the Shockley-Ramo theorem applied to RPC detectors, the 
induced signal is attenuated by a factor $A_{geom}=1 + 2 d / \epsilon_r g$, 
where $d$ is the electrode thickness, $g$ is the gas gap and $\epsilon_r$ is 
the bakelite relative dielectric constant.
However, in a wide range of $\epsilon_r$ values from 3.5 to 22,
less than a  10\% decrease of the attenuation factor for geometry 2 with
respect to geometry 1 is expected.
Therefore a comparison between the two considered geometries is basically
a comparison between the discharges in the two different gas gaps.

The tests described in this paper have been performed on RPCs operated in
streamer with gas mixtures made of Argon, TetraFluoroPropene (HFO-1234ze)
and Sulphur Hexafluoride (SF$_6$), shown in table \ref{tablegas}.
It is worth noticing that mixtures containing HFO-1234ze have been recently
proposed for use in RPCs (see for instance \cite{bGent2016} and \cite{bhfo}), 
both for avalanche and streamer mode operation, due to its low Global Warming 
Potential.

\begin{table}[h]
\centering
\caption{Summary of the gas mixtures used in the tests.}
\vspace{0.5cm}
\label{tablegas}
\begin{tabular}{|c|c|}
\hline
Mixture & Ar/HFO-1234ze/SF$_6$ volume ratios \\
\hline
 1 & 76.8/23.1/0.1 \\
 2 & 65.6/34.2/0.2 \\
 3 & 54.4/45.4/0.2 \\
 4 & 44.3/55.5/0.2 \\
\hline
\end{tabular}
\end{table}

\section{Results}
\label{sec:results}

In figure \ref{fig1} the efficiency is shown as a function of
{the operating voltage, for all the considered mixtures and geometries.
A comparison of all the efficiency plots for the 1 mm gas gap RPC shows
similar results as observed for 2 mm gap RPCs \cite{bGent2016}: the higher
the HFO concentration, the higher the operating voltage; low HFO
concentrations lead to higher multi-streamer probability, also shown in figure
\ref{fig1} as a function of the efficiency.
For 1 mm gas gap RPC the efficiency plateau value is around 90\%, lower
than for 2 mm gap RPC, independently from the HFO-1234ze concentration.

\begin{figure}
\centering
\includegraphics[width=\textwidth]{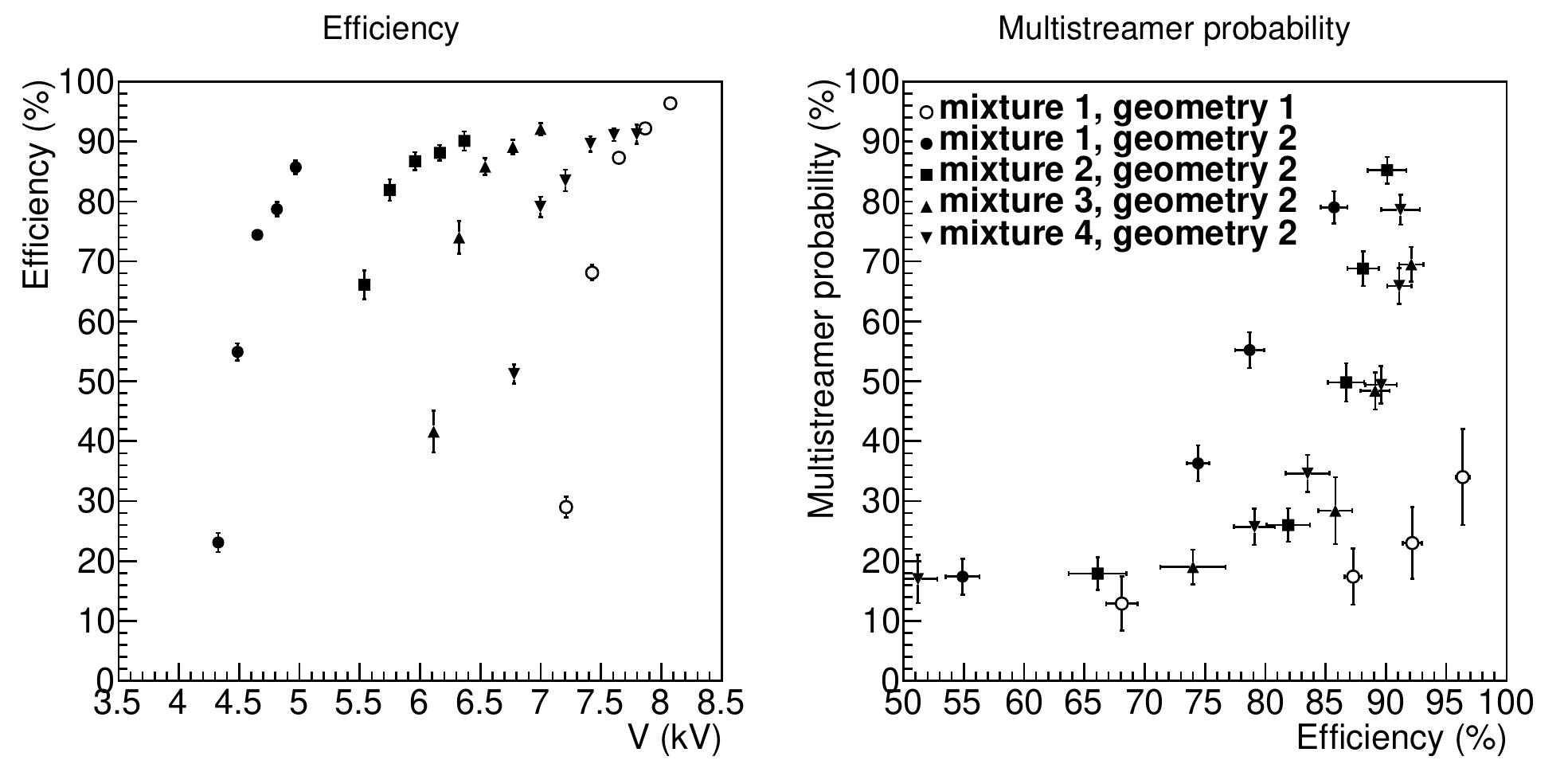}
\caption{Left plot: Efficiency as a function of the operating voltage for
different mixtures and gas gaps. Right plot: Multi-Streamer probability as
a function of the efficiency.}
\label{fig1}
\end{figure}

\begin{figure}
\centering
\includegraphics[width=10cm]{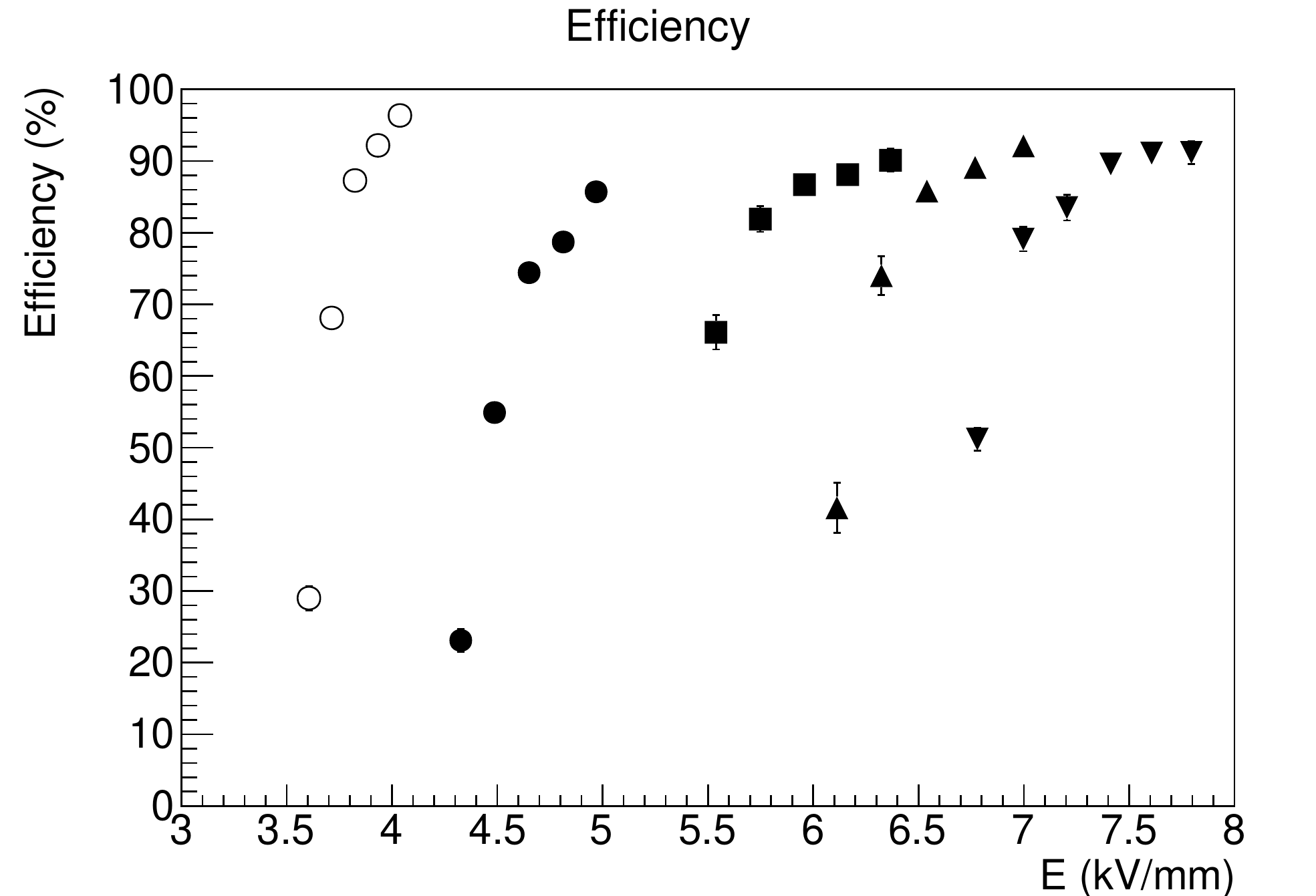}
\caption{Efficiency as a function of the electric field across the gas
gap for different mixtures and gas gaps. See figure \ref{fig1} for
label description.}
\label{fig2}
\end{figure}

A direct comparison between 1 mm and 2 mm gas gaps has been made using 
mixture 1: the efficiency plateau for 1 mm gap is reached at a much lower 
voltage with respect to the RPC with 2 mm gas gap.
In figure \ref{fig2} the efficiencies are shown as a function
of the nominal electric field, E, in the gas gap.
Assuming that the streamer is obtained when $\alpha(E) \times d \sim 20$ 
(Raether condition), where $\alpha(E)$ is the first Townsend coefficient 
and $d$ is the gas gap, decreasing the gas gap d, $\alpha$ must reach higher 
values (at higher electric fields) to operate the detector with good 
efficiency.    

Despite the greater easiness in streamer formation (higher efficiency plateau 
values and operation at lower electric field values),
the RPC with 2 mm gas gap shows a significant better quenching power (lower
multi-streamer probability) with respect to the one with 1 mm gas gap,
even for much lower concentrations of the quencher (HFO-1234ze) in the gas
mixture, as can be deduced from the multi-streamer probabilities shown in
figure \ref{fig1}. 
Assuming that multi-streamer events are due to UV photons, produced either in 
the avalanche precursor or in the following streamer, ionizing the gas or 
extracting an electron from the cathode, in 2 mm gas gap such photons have 
on average a larger path in the gas before hitting the cathode.
Therefore the plots of figure \ref{fig1} suggest that the production of 
multi-streamers is mainly due to electron extraction from the cathode rather 
than to gas ionization.

\begin{figure}
\centering
\includegraphics[width=\textwidth]{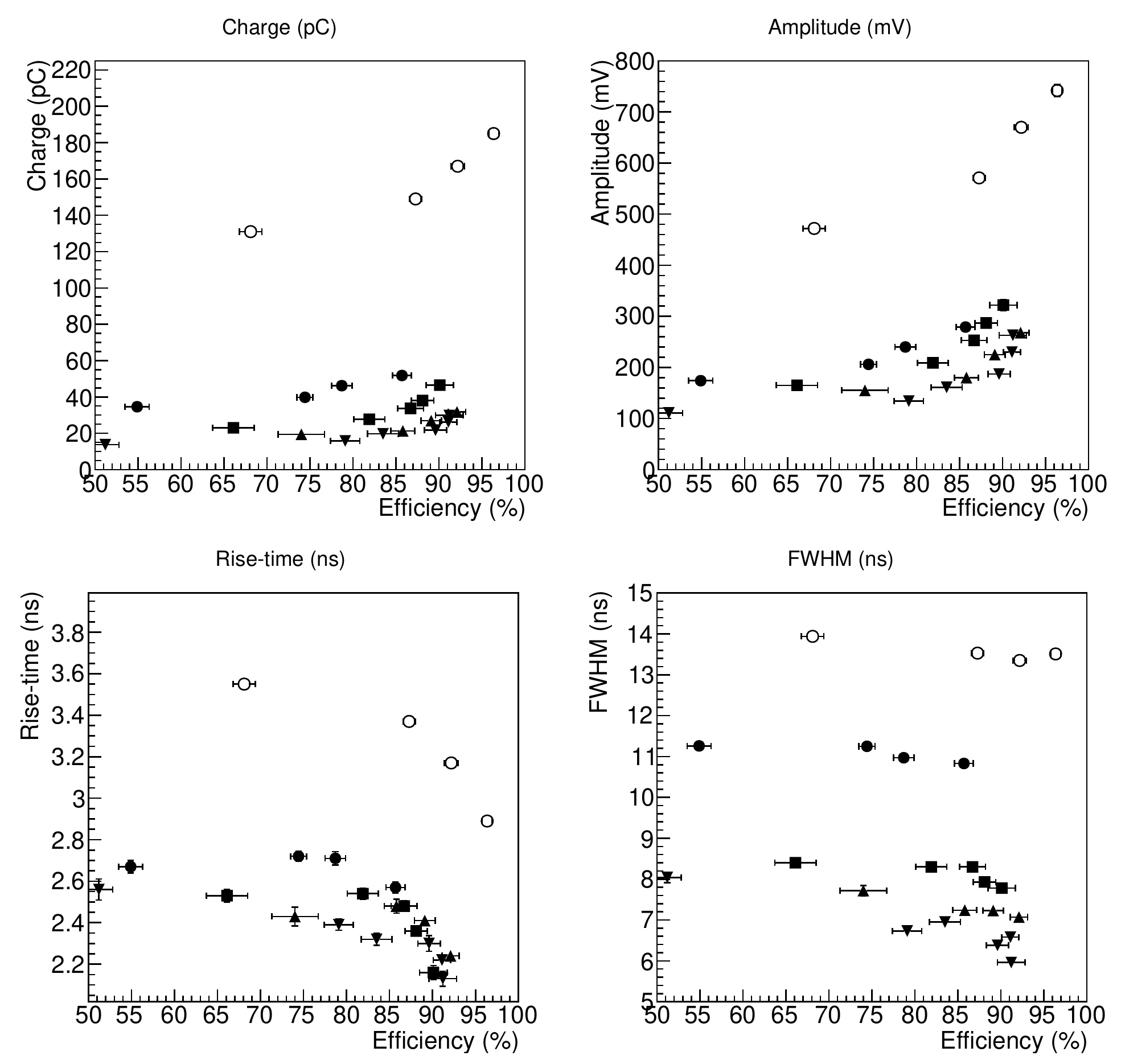}
\caption{Single streamer charge (top left), amplitude (top right), rise-time 
(bottom left) and Full Width Half Maximum (bottom right) as a function of the 
efficiency for different mixtures and gas gaps. 
See figure \ref{fig1} for label description.}
\label{fig3}
\end{figure}

In figure \ref{fig3} the single streamer parameters (total prompt
charge, signal amplitude, rise-time and FWHM) are shown as a function
of the efficiency.

Comparing data acquired on the RPC built according to geometry 2 with the 
considered gas mixtures, a decrease of the signal amplitude, FWHM and charge 
is observed for raising HFO-1234ze concentrations.
The signal amplitude for geometry 1 is between 2 and 3 times higher than
for geometry 2 at fixed efficiency values; the ratio is even higher for the 
prompt charge, due to the different signal FWHM.
The rise-time is slightly lower for 1 mm with respect to 2 mm gas gap, while 
an even small dependence is observed from the HFO-1234ze concentration, 
among the considered gas mixtures, in a wide range of electric field values
from 4 to 8 kV/mm. 

\begin{figure}
\centering
\includegraphics[width=12cm]{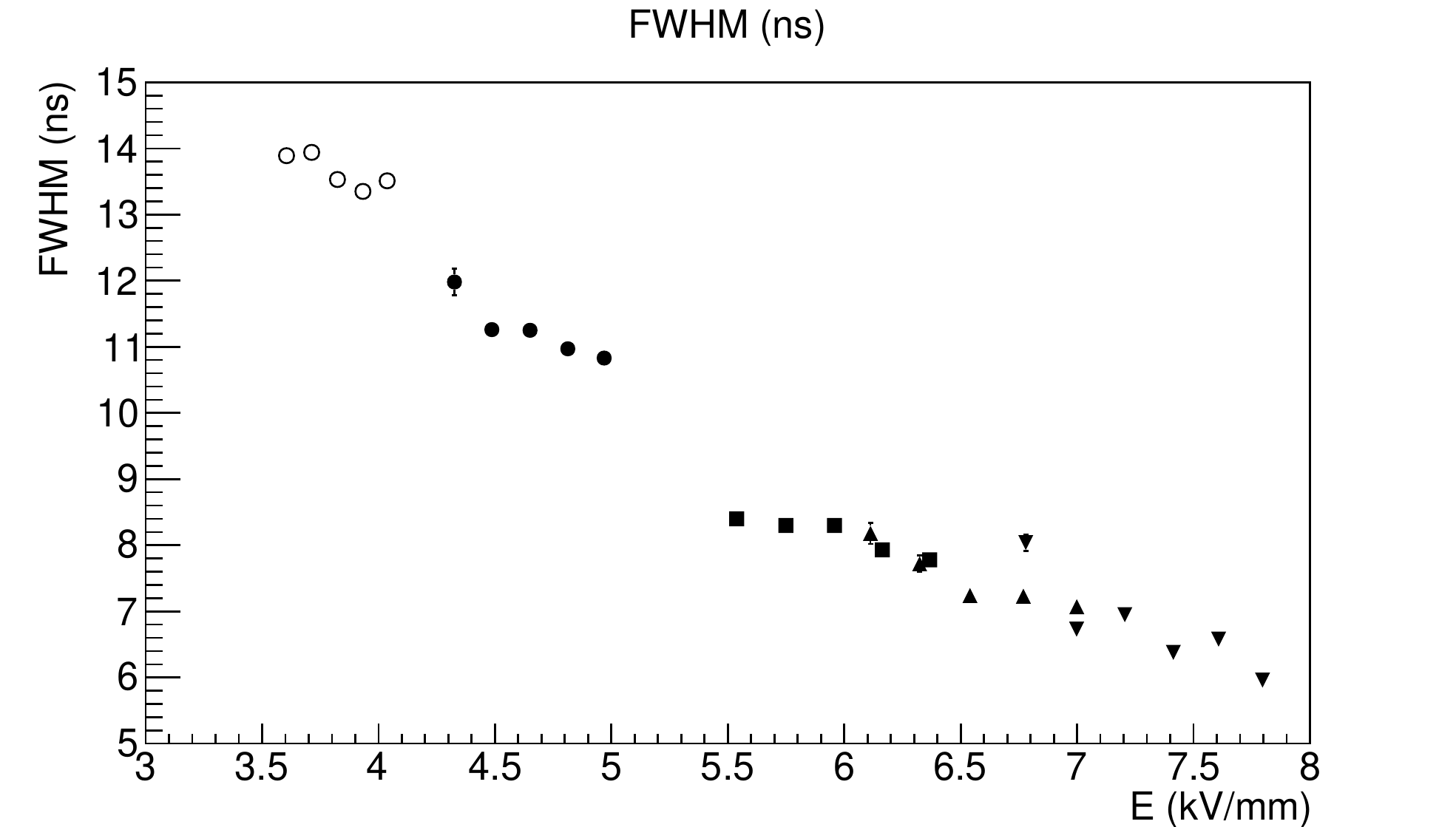}
\caption{Single streamer Full Width Half Maximum as a function of the 
nominal electric field across the gas gap for different mixtures and gas gaps. 
See figure \ref{fig1} for label description.}
\label{fig3b}
\end{figure}

By comparing results obtained both with different RPC geometries and with
the different considered gas mixtures, an almost linear
dependence is observed between the induced signal FWHM and the electric field,
about 2 ns every kV/mm, as shown in figure \ref{fig3b}.

Finally in figure \ref{fig4} the streamer formation time, measured with respect
to one of the trigger RPCs, used as a reference, and the time resolution
are shown.
Good time resolutions are obtained both with 1 mm and 2 mm gas gaps.
The increased gap value and the lower electric field delay the formation of
the streamer by about 5 ns in the 2 mm gas gap RPC with respect to the one
with 1 mm gas gap.

\begin{figure}
\centering
\includegraphics[width=\textwidth]{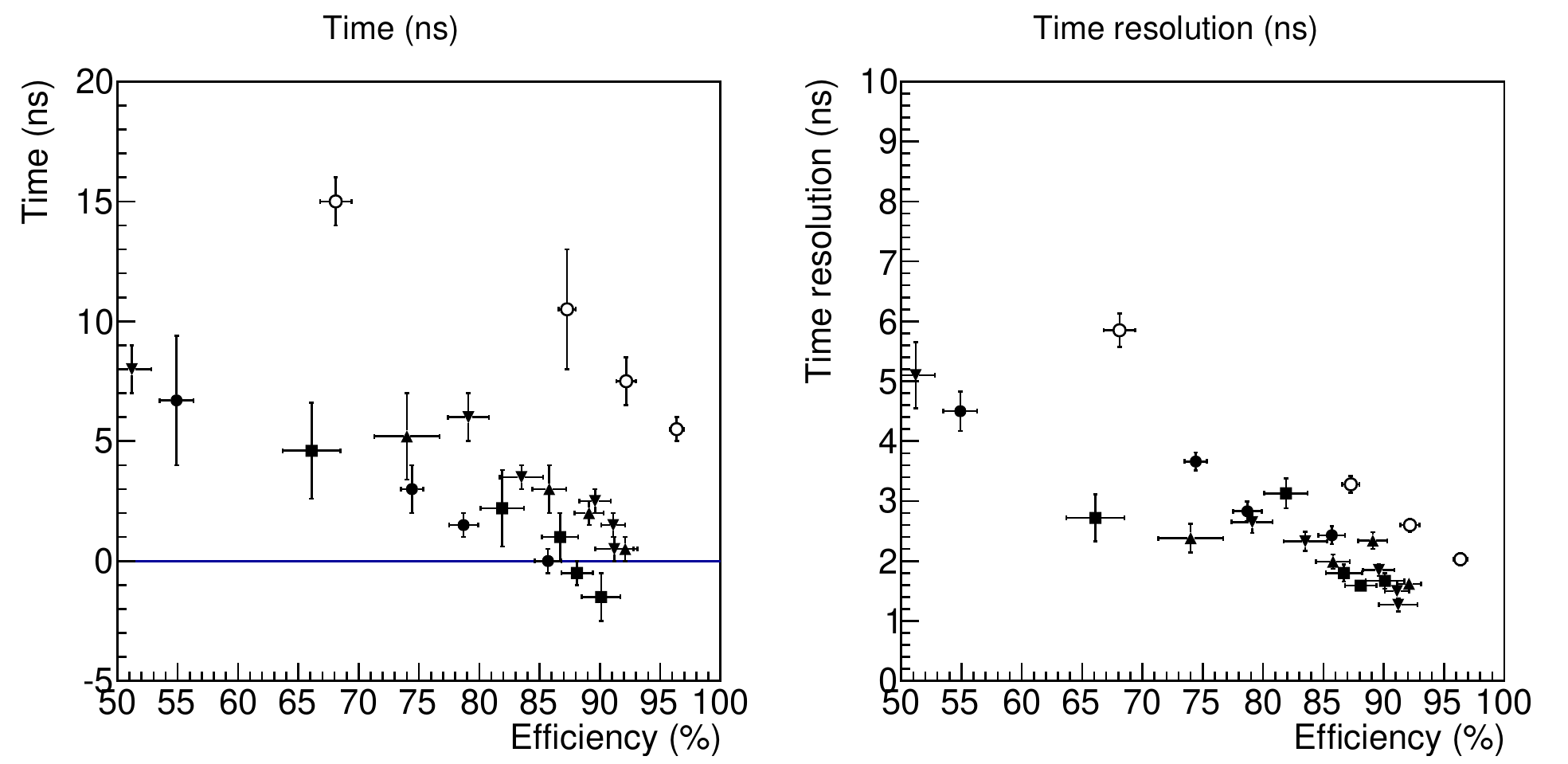}
\caption{Streamer arrival time (left) and time resolution (right) as a 
function of the efficiency for different mixtures and gas gaps. 
See figure \ref{fig1} for label description.}
\label{fig4}
\end{figure}

The average total charge ionized in the gas gap has been estimated, using
mixture 1 for the two considered detector geometries, from the ratio
between the operating current and the counting rate, after the subtraction
of ohmic contributions to the current.
The measured total charges, at the different operating voltages,
are comprised between 1 and 2 nC (0.5 and 1 nC) for 2 mm (1 mm) gas gap; they 
are not much different because of the higher multi-streamer probability with 1
mm gap compensating the lower charge.

\section{Conclusions}
\label{sec:conc}
In this paper a comparison is shown between the performances of RPCs operated 
in streamer with two different detector geometries, the old-fashioned 2 mm
gas gap and the newly proposed 1 mm gap; in both cases the electrode thickness
is similar to the gas gap, not to change the geometrical attenuation factor
of the signal induced on the read-out electrodes.

The 1 mm gap delays the formation of streamer to higher electric field
values, with a decrease of the plateau efficiency value to 90\%.
Streamer induced signals with 1 mm gap are smaller both in 
amplitude and width, but the multi-streamer probability is about as twice as 
that with 2 mm gap. 

Though in low rate applications multi-streamers could be not a problem,
the dramatic increase observed in streamer operated RPCs built according
to the new proposed geometry (1 mm thick gas gap and 1.2 mm thick bakelite 
electrodes) makes the old one (2 mm thick gas gap and 2 mm thick bakelite
electrodes) the preferred choice, at least with the mixtures used in the
tests reported here.


\end{document}